\documentclass[a4paper,fleqn]{cas-dc}

\usepackage[authoryear,longnamesfirst]{natbib}

\newcommand{\added}[1]{\textbf{#1}}
\renewcommand{\added}[1]{#1} %when submitting to preprint server or to provide clean LaTeX

\begin{document}
% \linenumbers % comment out with lineno package
\let\WriteBookmarks\relax
\def\floatpagepagefraction{1}
\def\textpagefraction{.001}

% Short title
\shorttitle{BCA Mineral Sputtering Limitations}    

% Short author
\shortauthors{Jäggi et al.}  

% Main title of the paper
\title [mode = title]{Limitations of Using BCA Codes for Modeling the Sputtering Behavior of Planetary Surfaces}  

% Title footnote mark
% eg: \tnotemark[1]
%\tnotemark[1] 

% Title footnote 1.
% eg: \tnotetext[1]{Title footnote text}
%\tnotetext[1]{} 

% First author
%
% Options: Use if required
% eg: \author[1,3]{Author Name}[type=editor,
%       style=chinese,
%       auid=000,
%       bioid=1,
%       prefix=Sir,
%       orcid=0000-0000-0000-0000,
%       facebook=<facebook id>,
%       twitter=<twitter id>,
%       linkedin=<linkedin id>,
%       gplus=<gplus id>]

\author[1,3]{Noah Jäggi}[orcid=0000-0002-2740-7965]

% Corresponding author indication
\cormark[1]

% Footnote of the first author
%\fnmark[1]

% Email id of the first author
\ead{noah.jaeggi@bluewin.ch}

% URL of the first author
\ead[url]{https://www.noahjaeggi.com}

% Credit authorship
% eg: \credit{Conceptualization of this study, Methodology, Software}
\credit{Conceptualization, Data curation, Formal analysis, Funding acquisition, Investigation, Methodology, Project administration, Software, Validation, Visualization, Writing (original draft), and Writing (review \& editing)}

% Address/affiliation
\affiliation[1]{organization={Laboratory for Astrophysics and Surface Physics, University of Virginia},
            addressline={395 McCormick Road}, 
            city={Charlottesville},
%          citysep={}, % Uncomment if no comma needed between city and postcode
            postcode={22904}, 
            state={VA},
            country={USA}}

\author[1]{Adam K. Woodson}[orcid=0000-0002-6112-7502]
%\fnmark[2]
\ead{akw8r@virginia.edu}
%\ead[url]{}
\credit{Conceptualization, Formal analysis, Investigation, Methodology, Software, and Writing (review \& editing)}

\author[2]{Paul S. Szabo}[orcid=0000-0002-7478-7999]

%\fnmark[2]
\ead{szabo@berkeley.edu}
%\ead[url]{}
\credit{Investigation, Writing (original draft), and Writing (review \& editing)}

% Address/affiliation
\affiliation[2]{organization={Space Sciences Laboratory, University of California},
            addressline={7 Gauss Way}, 
            city={Berkeley},
%          citysep={}, % Uncomment if no comma needed between city and postcode
            postcode={94720}, 
            state={CA},
            country={USA}}

\author[3]{Johannes Brötzner}[orcid=0000-0001-9999-9528]
%\fnmark[2]
\ead{broetzner@iap.tuwien.ac.at}
%\ead[url]{}
\credit{Investigation, Writing (original draft), and Writing (review \& editing)}

% Address/affiliation
\affiliation[3]{organization={Institute of Applied Physics, TU Wien},
            addressline={Wiedner Hauptstraße 8-10/E134}, 
            city={Vienna},
%          citysep={}, % Uncomment if no comma needed between city and postcode
            postcode={1040}, 
            state={},
            country={Austria}}

\author[3]{Friedrich Aumayr}[orcid=0000-0002-9788-0934]
\ead{aumayr@iap.tuwien.ac.at}
\credit{Supervision, Resources, and Writing (review \& editing)}

\author[1]{Catherine A. Dukes}[orcid=0000-0002-4549-7204]
\ead{cdukes@virginia.edu}
\credit{Project administration, Supervision, Resources, and Writing (review \& editing)}

%\ead[url]{}

% Corresponding author text
\cortext[1]{Corresponding author}

% Footnote text
%\fntext[1]{}

% For a title note without a number/mark
%\nonumnote{}

% Here goes the abstract
\begin{abstract}
%\textcolor{red}{WORD COUNT: 242/250}
Binary collision approximation (BCA) codes are potentially powerful tools to simulate ion irradiation ejecta properties, such as the composition and the angular and energy distributions of the sputter yield. However, recent advances in the sputtering of minerals have highlighted the low predictive fidelity of BCA codes such as SDTrimSP when compared to experimental measurements. We demonstrate how a sputtering model that underestimates the forward sputtering on a flat surface at large ion incidence angles from surface normal will lead to an erroneous result for rough and porous surfaces, where most ejected particles are directed along the surface normal. We demonstrate how this is the case for an existing model, which reliably predicts sputtering mass yields from a flat enstatite surface but fails to accurately reproduce the angular distribution of sputtered particles. We then compare this to a BCA model incorporating higher surface-binding energies---based on a molecular dynamics description of plagioclase---which underestimates mass yields but significantly reduces back-sputtering and better reproduces laboratory sputter angle distributions measured at large ion incidence angles. We conclude that the BCA model cannot simultaneously reproduce both the sputter yield and the sputter angle distribution arising from He irradiation of mineral targets, either due to the inherent geometric simplicity of the BCA or because the model neglects yield-enhancing processes such as molecule and cluster sputtering. This demonstrates a structural limitation of current BCA-based models when realistic surface morphologies are considered, rather than a problem that can be resolved by parameter tuning alone. 
\end{abstract}

% Use if graphical abstract is present
%\begin{graphicalabstract}
%\includegraphics{}
%\end{graphicalabstract}

% Research highlights
\begin{highlights}
\item When compared to experimental data, BCA codes like SDTrimSP can reproduce either the amount of sputtering or the direction of the ejected particles, but not both simultaneously. This mismatch reflects a fundamental limitation of the BCA approach rather than a tunable parameter issue.
\item While flat-surface simulations often agree with experiments, simulation results from rough and porous surfaces tilt the ejecta more toward the surface normal, and current models fail to reproduce this behavior accurately.
\item The use of higher, more realistic surface-binding energies produces better angular (tilt) distributions but drastically underestimates total sputter yields.
This underestimate could be explained through ejection of molecules and clusters, but BCA codes only handle single-atom sputtering, causing them to miss important yield-enhancing processes.
\end{highlights}

%\nocite{*}

% Keywords
% Each keyword is seperated by \sep

\maketitle

% ------ Main Section -------------------------------------

\section{Introduction}\label{sec:intro}

%Partial sputtering yields of Ta, TaO and O produced by 100--600~eV Ar ions impinging on Ta2O5 \cite{oechsner_sputtering_1978}. 
 
Understanding the formation of collisionless exospheres surrounding airless bodies in space is critical for the interpretation of in situ and ground-based exosphere observations. Several processes contribute to the exospheric particle populations, including thermal desorption, photodesorption, and electron-stimulated desorption; impact vaporization by micrometeoroids; and solar wind ion sputtering \citep[e.g.,][]{wurz_particles_2022}. To describe solar wind ion sputtering in exosphere models, the yield, composition, energy, and angular distributions must be known. The underlying sputtering model thereby informs on the distribution of ejected particles in the exosphere, where they are observed. The model thus ties the surface evolution and the exosphere observables together. In this work, we focus on the influence of macroscopic surface properties---such as topographic roughness and porosity---on the simulated angular distribution of ejecta predicted by current sputtering models %(Fig.~\ref{fig:1Dvs3D}) 
and on how well said distributions reproduce laboratory sputter yield and angular distribution data. %This is motivated by the fact that a rough surface would be representative of a rocky or rubble pile surface, whereas a porous surface is more representative of regolith, which is known to make up the majority of the surfaces of heavily weathered rocky bodies in space, such as the Moon.

\subsection{Angular distribution observations and theories}
For flat, polycrystalline surfaces, angular distributions of sputtered atoms produced by ions with normal incidence have been experimentally shown to follow cosine distributions with exponents exceeding unity. For example, the material sputtered from amorphous germanium by 80\,keV Ar$^+$ ions at normal incidence approximately follows a $\text{cos}^{1.57}\theta$ distribution \citep{andersen_angular_1985}. These results are comparable to those of other polycrystalline metals \citep{Andersen1982,Andersen1983}. This finding led to the assumption that a power-of-two angular dependence best describes flat, polycrystalline surfaces \citep{Hofer1991}. For airless planetary bodies, porous regolith surfaces are relevant and a simple cosine distribution was proposed by \cite{Cassidy2005}, which was consequently implemented in exosphere modeling works \citep[e.g.,][]{Wurz2010}. Recently, improved experiments and new computational models of the sputtering of rough and porous surfaces allow for further testing of this underlying assumption. For this purpose, we compare ejecta distributions obtained from simulated flat, rough, and porous surfaces with available experimental data.

%In the icy moon exosphere model by \cite{wieser_emission_2016}, the authors instead assumed a sub-quadratic, non-isotropic angular distribution of the emitted particles proportional to $\text{cos}^{4/3}\theta$, where $\theta$ is the angle to the surface normal, assuming that the emission angle of the sputtered particles is independent of the angle of incidence of the projectile based on laboratory work by \citep{vidal_angular_2005}. In this work we review the angular distribution from the major rock forming pyroxene MgSiO$_3$ when simulated for either a flat, rough, or porous surface.

\subsection{Flat surface ejecta distributions}
Under normal ion incidence, there appears to be good agreement between measurement and simulation of the angular distribution of ejecta from flat surfaces. For example, the distribution of sputter ejecta induced by 4\,keV\,He$^+$ normally incident on enstatite (MgSiO$_3$), as simulated with the binary collision Monte Carlo code SDTrimSP \citep{Mutzke2024}, can be fit with a cosine distribution having exponents exceeding unity ($\mathcal{H}(\theta) := \cos^{1.7}(\theta)$ for $\theta<\mathrm\theta_\text{pl}$  and  $\cos^{1.2}(\theta)$ for $\theta>\theta_\text{pl}$) but below 2 \citep{jaggi_new_2023,Jaggi2024}. This agrees well with the 1.57 exponent found in \cite{andersen_angular_1985}.
An elongated sputtered plume with $\text{cos}^{>1}$ is also seen in laboratory sputtering data from a flat MgSiO$_3$ thin amorphous film \citep{Biber2022}. There, irradiation was done at oblique incidence angles ($\alpha_\text{in}$) of 45$^\circ$ and 60$^\circ$ with respect to the sample surface normal. Laboratory data have shown a significant forward tilt $\geq45^\circ$ away from the incident ion beam at large $\alpha_\text{in}$ from the surface normal due to near-surface momentum transfer in single-scattering events. This is unlike the simulated angular distributions that were performed for the same surface in \cite{jaggi_new_2023}. Underestimating the enhanced forward tilt of ejected particles may be relevant when researching the exospheric distribution and amount of atmospheric loss of a rocky body in space because grazing trajectories are more likely to redeposit nearby.%, while they can also receive a boost in tangential velocity due to planetary rotation, which would then increase the likelihood of escape or lengthen the particle's orbital period. Whether this effect is noticeable on the exosphere scale is unclear, nevertheless, 
In any case, the ejecta tilt gives us another observable to validate fundamental model parameters, such as the binding energy, which is still an unresolved area of research.

One way to increase the forward tilt of sputtered atoms in simulation is by increasing the contribution of atoms ejected from the first atomic layer relative to the contributions of the deeper atomic layers \citep[discussed in][]{jaggi_new_2023}. Atoms ejected from the first layer have a narrow distribution with a forward-faced component which is informed by the ion impact angle, whereas atoms originating from the second atomic layer or beyond form broader angular distributions centered nearer to the surface normal unless channeled in some preferential direction by the crystal structure \citep{Schwebel1987,Whitaker1993,gnaser_low-energy_1999}. In a binary collision simulation, this can be achieved in two ways. On one hand, one can increase the planar surface potential that must be overcome for an atom to leave the surface \citep{Eckstein1991,robinson_theoretical_1981}, i.e. the surface binding energy (SBE). On the other hand, one can increase the energy required to break bonds in the target, referred to as the bulk binding energy (BBE). An increased SBE thereby prevents low-energy recoils created late in a collision cascade from escaping the target, whereas an increased BBE rapidly depletes the energy of the impinging ion within the collision cascade, increasing the number of subsequent low-energy recoils. 

\begin{figure}[htb!]
    \centering
    \includegraphics[width=\linewidth]{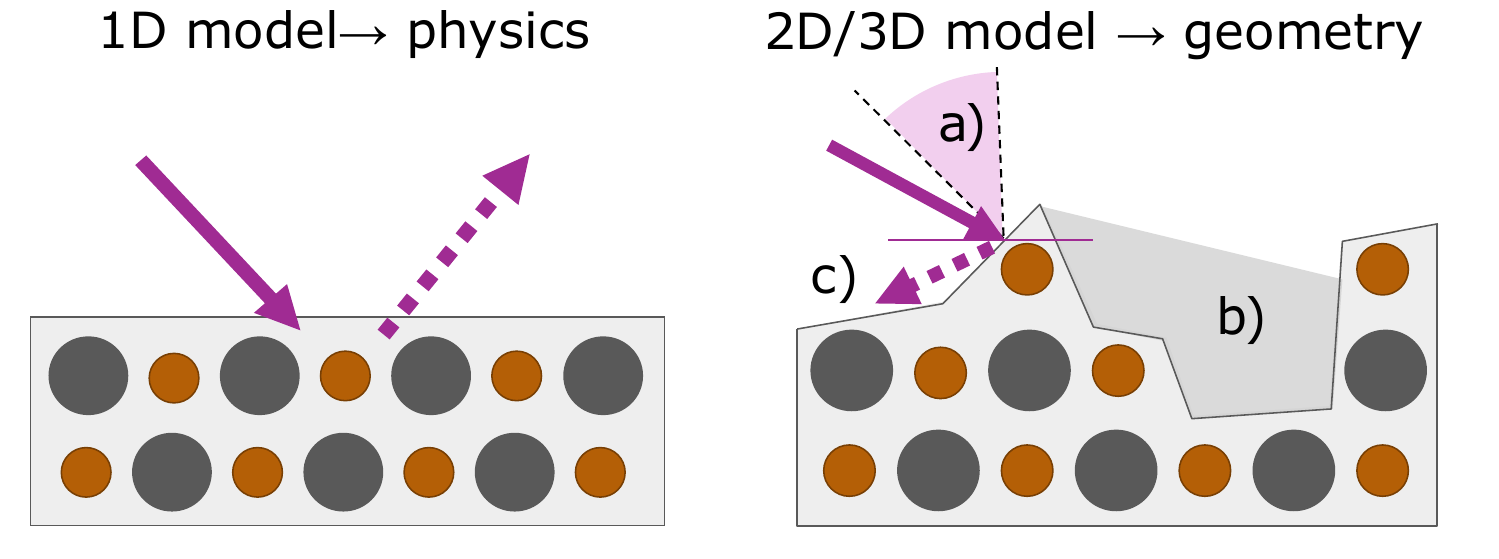}
    \caption{To evaluate a sputtering model, flat surfaces allow for mitigating geometrical effects, revealing the underlying physics (left). On a rough/porous surface, which is more representative of soils from airless rocky bodies, such as the Moon, the sputtering behavior is affected by a) variations in local incidence angles, b) shadowing, and c) re-deposition of material.}
    \label{fig:1Dvs3D}
\end{figure}

The angular distribution tilt of the ejecta is expected to be heavily affected by the target surface morphology, including surface roughness and/or porosity This was observed in work on tungsten \citep[Figs.\,15\,\&\,16 in ][]{Stadlmayr2020} and copper \citep{bu_absolute_2024}. Because mineral surfaces are difficult to produce in an ideal, flat manner - either due to mineral pelletization (inherently rough), difficulties in obtaining large single crystal specimens, or because polishing is imperfect for polycrystalline materials - experimental results for angle-resolved sputter distributions are often described by sputter physics convolved with inherent geometrical effects due to surface topography (Fig.~\ref{fig:1Dvs3D}). Here, we demonstrate to what degree established sputtering models can reproduce experimental distributions from flat thin films and rough, pressed pellets.

%We will discuss the limitations of the increased binding energies in reproducing experimental data by applying the latest silicate binding energy data published in \cite{morrissey_solar_2024}. 

\subsection{Rough and porous ejecta distributions}
\added{For a porous, regolith-like surface, no detailed experimental measurements of sputtering distributions exist. Preliminary measurements involving loose, 53-90\,$\mu$m Cu grains irradiated by 20~keV Kr$^+$ indicate that backward-directed (toward the incident ion source) sputter yields significantly exceed those in the forward direction \citep{bu_absolute_2025}. This behavior was subsequently reproduced by using a multiscale MD-BCA-MC approach \citep{verkercke_theoretical_2026}. The input for the angular distribution of the sputtered particles is thereby sourced from SDTrimSP, with the simulated Cu ejecta, using default SBEs (enthalpy of sublimation), reproducing the flat surface laboratory data well \citep{bu_absolute_2024}.} Solar wind protons scattered from the porous regolith surface of the Moon were also observed to recoil preferentially back toward the source \citep{schaufelberger_scattering_2011}, which was initially supported by simple 3D BCA simulations \citep{szabo_energetic_2023} \added{followed by more computationally heavy MD simulations which captured the full extent of backscattering \citep{verkercke_effects_2023}}. 

\added{Unlike the results of backscattered ions and Cu sputtering ejecta from Kr ions, \cite{Biber2022} performed laboratory experiments with rough, pressed enstatite pellet surfaces and found that the forward tilt from the surface normal was still significant at 21–45$^\circ$. This observation thereby agrees with that found for an Apollo 16 regolith pellet shown in \cite{brotzner_sputter_2025} \citep[preprint of ][]{brotzner_solar_2025}.} And although \cite{Biber2022} were able to explain the difference in absolute mass yields between flat and rough surfaces by using a rough surface model \citep[SPRAY;][]{Cupak2021}, the work did not compare simulated rough surface \textit{angular distributions} to experimental data. In this work, we extend these studies by generating and characterizing angular distributions for rough and  porous surfaces. We reproduce the flat and rough MgSiO$_3$ simulations discussed above, apply the latest proposed mineral-specific binding energies \citep{morrissey_solar_2024}, discuss the resulting angular distributions in relation to previous works\citep{Biber2022,jaggi_new_2023}, and produce new data for porous surfaces following \cite{Szabo2022c} for comparison. 

While angular distributions from flat surfaces are relatively well understood, significant uncertainties remain for rough and porous surfaces, particularly in reproducing forward tilts of experimental data and accounting for potential simulation artifacts. This work addresses these gaps by systematically simulating sputter ejecta from flat, rough, and porous MgSiO$_3$ surfaces using SDTrimSP, SPRAY, and SDTrimSP-3D. We analyze the resulting angular distributions using a novel three-dimensional plume-fitting approach to evaluate the influence of surface morphology and binding energy models on sputter behavior. \textbf{We will show that unlike the Cu ejecta sputtered by 20 keV Kr ions, angular distributions from enstatite are not well reproduced in SDTrimSP.}

% \begin{figure*}[htbp!]
%     \centering
%     \includegraphics[width=1\textwidth]{archive/He_En_sbbx_0_20_45_60_72_80_adist_phi.pdf} 
%     \caption{\njnote{will be removed} Polar angular distribution of normalized mass yield $Y_\text{norm}$ sputtered from MgSiO$_3$ by 4~keV He$^+$ at increasing incident angles ($\alpha_{in}$) compared to experimental data \cite[black; ][]{Biber2022} with 2 standard-deviation errors.}
%     \label{fig:adist_phi}
% \end{figure*}

\section{Methodology}

To understand the physical response of a surface under ion irradiation, it is recommended to work with flat, amorphous surfaces to rule out geometry effects on the sputtering process. The latest advances in our understanding of solar wind ion sputtering from flat mineral surfaces are summarized in \citep{Biber2022,wurz_particles_2022,morrissey_establishing_2023,jaggi_new_2023}. To obtain sputter yields from realistic regolith surfaces, geometric effects must be considered (Fig.~\ref{fig:1Dvs3D}). The simulated data presented here were produced for flat, rough, and porous surfaces using the SDTrimSP \citep{Mutzke2024}, SPRAY \citep{Cupak2021,broetzner_spray_2025}, and SDTrimSP-3D \citep{VonToussaint2017} numerical codes, respectively. The parameters chosen for binding energies and target densities follow the ones used in \cite{Jaggi2024}, which were found to best reproduce the polar angular distribution of ejecta measured in \cite{Biber2022}.

To directly compare with the experimental data obtained in \cite{Biber2022}, we ran our simulations dynamically, where the layer compositions are allowed to change with fluence, except for the SDTrimSP-3D cases, where it was not possible. Experimental measurements of raw mass yield vs. fluence demonstrate that total mass yield is typically maximized at the onset of irradiation as adsorbate species such as water or adventitious carbon are quickly removed, after which it decreases to the yield of the surface in equilibrium with the incident irradiation \citep[e.g.,][]{Biber2020,Szabo2020b}. This equilibrium yield occurs as the surface elemental concentrations reach a steady state based on the preferential removal of low-mass, low-binding energy species with ion implantation. Implantation of ions has been neglected in this work due to insufficient experimental data and the negligible effect on equilibrium yields reported by \citep{Biber2020}.

\begin{figure*}
    \centering
    \includegraphics{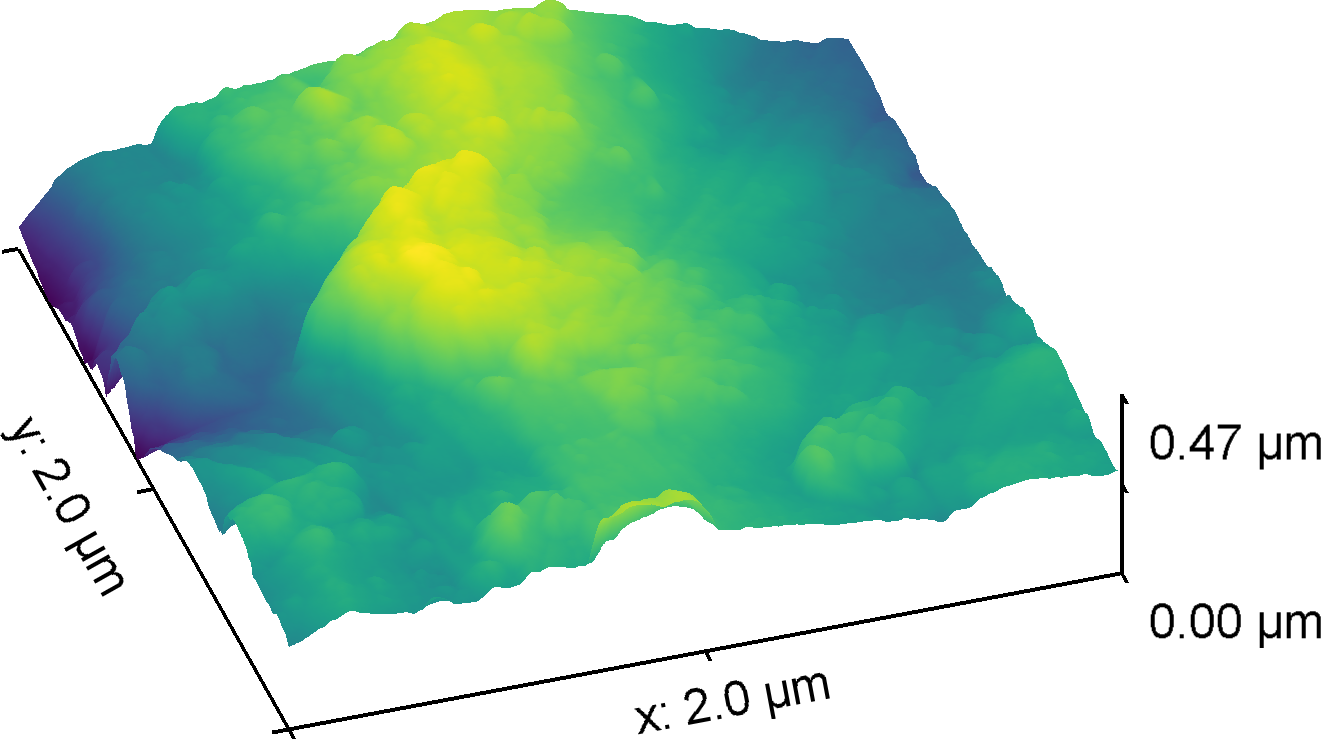}
    \includegraphics{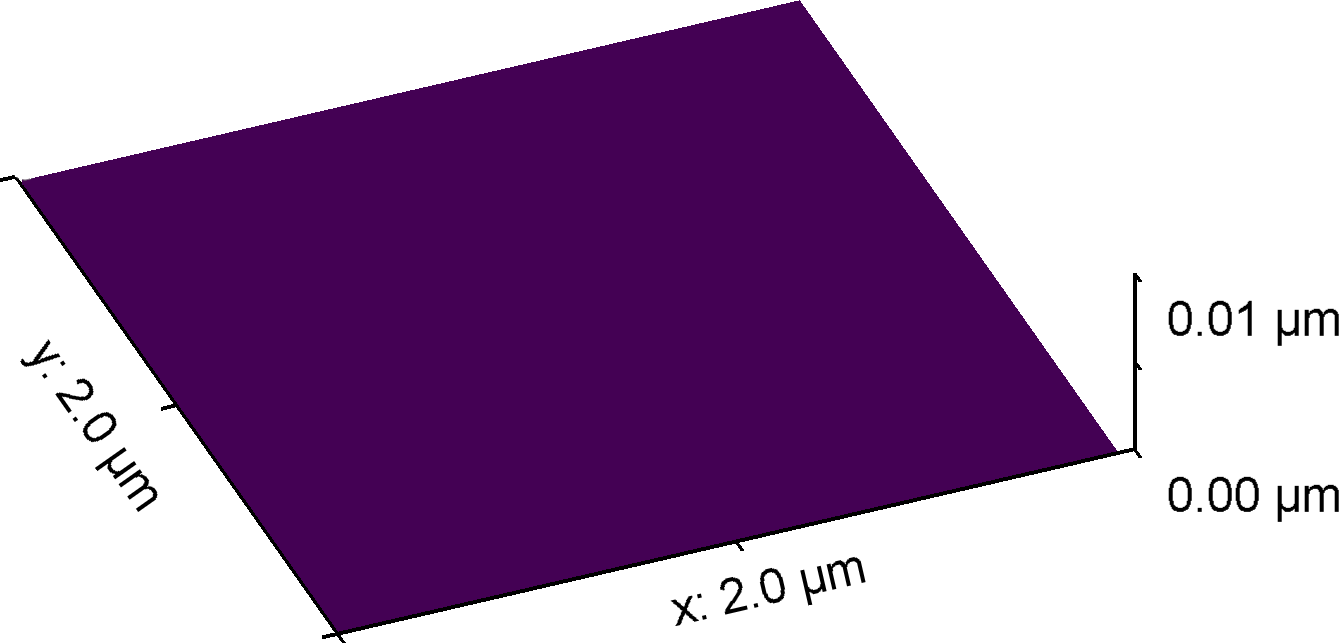}
    \caption{Surface structures used in SPRAY}
    \label{fig:structures}
\end{figure*}

\subsection{SDTrimSP: flat surface}\label{meth:SDTrimSP}

The flat surface data considered here are the same as in \cite{Jaggi2024}. The MgSiO$_3$ flat target is simulated by using MgO and SiO$_2$ as components, which results in a more accurate initial description of the mineral density \citep[ideal: 3.20~g\,cm$^{-2}$, compound model: 3.05~g\,cm$^{-2}$, default atomic densities: 1.74~g\,cm$^{-2}$, in][]{jaggi_new_2023}. The binding energy model ``isbv=4'' that was applied here uses both the default SDTrimSP surface binding energies (enthalpy of sublimation/dissociation of elements) combined with bulk binding energies given by the enthalpy of formation of MgO and SiO$_2$. We refer the reader to \cite{Jaggi2024} for an in-depth discussion about the model and the underlying assumptions.%, and how the results compare to SRIM data. Regarding the latter, it was already shown for incidence angles $<60^\circ$ that SRIM shows the largest deviation in angular distribution of yields from experimental data when compared to either SDTrimSP and TRIDYN \citep{Hofsass2014}, and no SRIM data was used as a result.
\added{We note here, that the degree of how damage changes binding energies is limited in the model by the transition from elements bound to a compound to atomic elements that have binding energies that are equal to the default enthalpy of sublimation.}
To improve statistics \citep[compared to][]{Jaggi2024}, the dynamic SDTrimSP simulations were run twice: first until the surface-irradiation equilibrium was reached, and then a second time using the pre-equilibrated surfaces. This provided plume data that excluded pre-equilibrium trajectories, which must be isolated in this way because SDTrimSP does not report the fluence at which a given trajectory is recorded.

\subsection{SPRAY: flat and rough pellet surface}\label{meth:SPRAY}

We used SPRAY to model the sputtering of rough surfaces from pressed MgSiO$_3$ pellets as an analog to \cite{Biber2022}, where it was demonstrated that the SPRAY code accurately reproduces the reduced yield from a rough surface as compared to a flat surface. These samples were prepared as described in \cite{jaggi_creation_2021} and their surfaces were characterized by atomic force microscopy (AFM). 

The physical input for SPRAY simulations is a data repository generated from one-dimensional BCA simulations. For this study, we used SDTrimSP with ``isbv=4'' binding model as described in Section~\ref{meth:SDTrimSP} as inputs; thus, for flat surfaces, the same yields, energy, and angular distributions are recovered from SPRAY as in the SDTrimSP flat surface case. The simulations were run dynamically using pre-equilibrated surfaces to maximize efficiency in obtaining trajectories from the equilibrated surface. SPRAY maps these flat surface results onto AFM images of real sample surfaces (Fig.~\ref{fig:structures}) via a ray-tracing algorithm: For a given impact position on the triangulated surface, the local incidence angle of the impacting ion is evaluated and the trajectories of reflected and sputtered atoms are statistically generated based on flat surface angular distributions precomputed using SDTrimSP. Ejected particle trajectories are subsequently traced until they either reach the vacuum or intersect a surface triangle. For a more detailed description of the code, the reader is referred to \cite{broetzner_spray_2025}. Once the data repository is created, SPRAY is typically computationally less demanding than SDTrimSP-3D for extended surfaces of physical samples but is limited to non-porous morphologies as resolved by AFM. 

\subsection{SDTrimSP-3D: porous regolith surface} \label{meth:SDTrimSP-3D}

Regolith analogs in the SDTrimSP-3D code were simulated using a porous stacking of \added{100} equally sized spheres, equivalent to the approach used in \cite{Szabo2022c, szabo_energetic_2023,szabo_updated_2025}. In this approach, a porous grain pile is built by dropping semi-spherical grains (limited by voxel resolution) into a box, subject to a sticking coefficient.  A porosity of around 80\% is allowed, which is necessary to achieve agreement with Chandrayaan-1 observations of solar wind protons scattering from the lunar regolith as neutral H atoms \citep{Szabo2022c}. For each case, simulations with five different randomly generated grain stackings are performed \added{since results dependent on the local structure \citep{szabo_energetic_2023}}. Due to computational limitations, it is not possible to simulate realistic micron-scale (or larger) grains in SDTrimSP-3D, but previous research has shown that grain-size-related effects become negligible if the grain radius is at least five times larger than the ion-implantation range \citep{nietiadi_sputtering_2014, Szabo2022c}. \added{More specifically, \cite{nietiadi_sputtering_2014} show that the ion interactions with spherical nanoparticles become size-independent if the grain is large compared to the size of the collision cascade. This is clearly the case for lunar regolith grains with sizes on the order of tens of $\mu$m and ions with implantation ranges of tens of nm.} For the case of 4\,keV~He ions impacting the enstatite surface studied here, SDTrimSP-3D predicts a mean He implantation range of 33\,nm. We therefore chose grain radii of 400\,nm for our simulations—40 times larger than the SDTrimSP-3D voxel resolution of 10\,nm$^3$ employed in these calculations. These settings thus fulfill the requirements of large enough grains while also ensuring the requirement of a resolution that is finer than the dimensions of the collision cascade \citep{VonToussaint2017}. \added{Lastly, we do not consider the effects of grain size distributions in this study, but the simulation cases with distributions by \cite{verkercke_effects_2023} indicate that the behavior of the distributions is similar to regoliths with uniform grain sizes. Thus we consider it to be a minor approximation here for the ion-regolith interaction to assume uniform grain sizes. }

In SDTrimSP-3D, the implementation of a binding energy model equivalent to ``isbv=4'' in SDTrimSP was not available. To approximate ``isbv=4'' in SDTrimSP-3D, we prescribed the density, SBEs, and BBEs of the SDTrimSP model to the SDTrimSP-3D model. The 3D model therefore behaves like the ``compounds without dissociation or recombination'' in \cite{Jaggi2024}, where bulk binding energies are derived from oxide compounds but without the possibility for said compounds to break down and reduce the bulk binding energy in the altered surface. When interpreting the resulting sputter data, we would thus expect overall lower yields, higher characteristic energies of sputtered atoms, and the flat-surface ejecta distribution trending in the forward direction, away from the surface normal and away from the direction of incidence.

\subsection{Increased binding energy cases}\label{meth:morrissey}
In SDTrimSP, the ejecta that are non-normal to the irradiated surface are derived from single collision recoil events (SDTrimSP output parameter gen/coll = 1). This suggests that adjusting the SBE is the key parameter to affect the tilt angle of the ejecta plume. Researchers are still actively calculating and measuring surface and bulk binding energies for silicates, but the current consensus is that the enthalpy of sublimation alone, as a proxy for SBE, is incapable of reproducing experimental data \citep[discussed in detail in][]{jaggi_new_2023}. For an educated guess regarding the increased binding energies in MgSiO$_3$, we have used element-specific binding energies for plagioclase that were computed via molecular dynamics \citep[Table~5: Elemental Best-fit Values $E_b$ in][]{morrissey_solar_2024}. Mg, however, does not occur in plagioclase and no computed values exist at the time of this writing, so we estimated a six-fold increase in binding energy relative to the enthalpy of sublimation value tabulated in SDTrimSP. This is in line with the 5.57 times increase for the best fit $E_b$ of Si, 4.3 for O, 6.5 for Al, and 7.4 for Na as reported \added{for the (001) surface of albite} in \cite{morrissey_solar_2024}. \added{This results in elevated SBEs for entatite of $E_{b, \text{Si}} = 26.3 $\,eV, $E_{b, \text{O}} = 10.4 $\,eV, and $E_{b, \text{Mg}} = 9.06 $\,eV.}

There is one caveat, however: \added{the SBEs reported in \cite{morrissey_solar_2024} depend on the specific surface site and increases with depth due to increased coordination between atoms. The SBEs would therefore follow a distribution instead of a single value, which we do not reproduce in the simulation. We instead use upper estimates for the SBEs to representing a case that may or may not be an endmember case, depending on the as of yet unknown, actual SBEs of enstatite. }

\subsection{Primary knock-ons in simulations}
The SDTrimSP sputter model used in \cite{jaggi_new_2023,Jaggi2024} exhibits a component that is made up of primary-knock-on recoils that form a sharp feature that becomes increasingly apparent in the ejecta angular distribution at large $\alpha_\text{in}$ ($\geq60^\circ$) when collapsed onto the polar plane. In the \cite{Biber2020} laboratory angular sputtering yield data, the data for 4\,keV He  at $60^\circ$ incidence for a flat MgSiO3 surface also shows a more forward-facing component for ejected atoms (Fig.\,\ref{fig:FLATsbbxVSsbxmYandDist} in Sec.~\ref{res:incSBEs}), but whether this represents a single collision peak is still unknown. For forward-facing, scattered primary ions, it has been shown that a characteristic single collision peak exists in the energy distribution of simulated particles but is absent in the laboratory data \citep[14\,keV Sn$^+$ on Mo;][]{Wilhelm2023}. On the contrary, recent experiments involving ruthenium irradiated with 15\,keV Ar$^{2+}$, Kr$^{2+}$, and Xe$^{2+}$ have demonstrated the presence of peaks in the energy distributions of scattered ions and single-collision sputtered particles % \njnote{Fig. 4 in Assink - are the green shaded areas actually where sputtered Ru atoms would occur?} \jbnote{Not really... The green shaded areas are those where \emph{recoils} would reside. In the paper, these are defined as being knocked out by an incident ion \emph{directly}, by one single ion-atom scattering event. In that case, their resulting energy is given by classic scattering kinematics and depends on the scattering angle, i.e., the detector position. If particles leave the solid \emph{sputtered}, i.e., after a lot of the incoming ion information is lost in the collision cascade, they have much lower energies.} 
depending on the incident ion \citep{Assink2024}. It is therefore difficult to conclude whether this feature—so prominent in SDTrimSP flat-surface simulations—is an artifact of the simulation or derives from an unidentified physical process. In this work we only attempt to capture the feature in our data-fitting procedure to understand its origin. However, we do not discuss its relative contribution to sputter yields derived from rocky, airless bodies.

\subsection{QCM detection filter for 2D data}
The data published in \cite{Biber2020} represents a slice of the ejecta captured by a quartz-crystal microbalance (QCM), positioned 17\,mm from the target , with a mass sensitivity that falls to 1/2 max at 1 mm from the center of its sensitive area \citep{Szabo2017}. Neither of the \cite{jaggi_new_2023,Jaggi2024} studies filtered the simulated data according to QCM solid angle when comparing it to experimental data, which led to an underestimation of the single collision peak intensity and the plume tilt angle. We find that when collapsing all trajectories onto the polar plane, instead of only projecting the QCM-detected center slice of the plume, the trajectories are more normally distributed.  When compared to experimental values, the data presented in Section~\ref{sec:results} Results is thus correctly filtered for comparison with the QCM measurements and the plot is labeled with ``QCM detection'' to emphasize this (Figs.~\ref{fig:FLATsbbxVSsbxmYandDist}\,\&\,\ref{fig:ROUGHsbxmDist} in Secs.~\ref{sec:results}).

\subsection{Data Fitting}\label{meth:fitting}
In the introduction, we established that there have been simple cosine distributions deployed in exospheric modeling to simulate the angular distribution of sputter products. We challenge this assumption based on the non-normal tilts observed in the angular distribution data of \cite{Biber2022} and the regolith work by \cite{Szabo2022c}. Furthermore, the azimuthal angle dependence of the ejecta had hitherto been neglected by exospheric modelers due to a lack of experimental data on rock-forming minerals.% \citep[work on Cu exists][]
For a complete implementation of experimental sputter findings, the yield has to be described in three dimensions. In SpuBase—a tool to provide baseline solar wind sputter yields from minerals and rocks for the modeling community—the azimuthal angle was not fit, which hampers adaptation \citep{Jaggi2024}. We propose that it is more accurate to fit the entire angular distribution of sputtered species with a three-dimensional plume, which is now implemented in SpuBase-v2.0 (doi: \href{https://doi.org/10.5281/zenodo.18741496}{10.5281/zenodo.18741496}). %and use the fitting results to discuss the capabilities of sputter codes and highlight uncertainties and open questions.

The laboratory data plumes were fit using 1) a single tilted lobe or 2) the sum of a tilted center lobe and two symmetric tilted side lobes (tri-lobe). Each lobe is characterized by a polar tilt angle, $\theta_\text{tilt}$, and an azimuthal rotation, $\phi_\text{tilt}$, which together \added{define the transformation $T(\theta_\text{tilt}$, $\phi_\text{tilt})$} that establishes the orientation of the lobe coordinate axes $\hat{x}_{lb}$, $\hat{y}_{lb}$, and $\hat{z}_{lb}$ with respect to the simulation coordinate axes. The model is formulated to transition smoothly to zero at $\theta_\text{lb}=\pi/2$  (i.e., perpendicular to $\hat{z}_{lb}$) and to avoid negative values. Furthermore, the lobe model includes two variable exponents ($m$ and $n$) that are used to stretch the lobe along the $\hat{x}_{lb}$ and $\hat{y}_{lb}$ axes. \added{In the lobe coordinate system, a single lobe is defined as follows}:
\begin{equation}
\label{eq:lobe}
    \rho_\text{lb} = A\left(\frac{1 + \cos(2 \theta_\text{lb})} {2}\right)^{n\,\cos^2(\phi_\text{lb}) + m\,\sin^2(\phi_\text{lb})},
\end{equation}
\noindent with
\begin{equation}
\begin{aligned}
    \theta_\text{lb} &= \arccos(z_\text{lb})\\
    \phi_\text{lb} &= \arccos\left(\frac{x_\text{lb}}{\sqrt{{x_\text{lb}}^2 + {y_\text{lb}}^2}}\right)
\end{aligned}
\end{equation}
where $A$ is the lobe magnitude along the $\hat{z}_{lb}$ axis and ($x_\text{lb}$, $y_\text{lb}$, $z_\text{lb}$) denotes a point on the lobe surface. \added{After coordinate transformation of each lobe to the simulation system, which embeds the tilt angles into the lobe formula, the full plume is obtained by summing lobes in simulation coordinates:}
\begin{equation}
\begin{aligned}
\label{eq:tri-lobe}
   \rho_\text{tot} = \rho_\text{sim}(\theta_\text{tilt,1}, \phi_\text{tilt,1}=0, A_1, m_1, n_1)\\ + \rho_\text{sim}(\theta_\text{tilt,2}, +\phi_\text{tilt,2},  A_2, m_2, n_2)\\ + \rho_\text{sim}(\theta_\text{tilt,2},-\phi_\text{tilt,2}, A_2, m_2, n_2)
   \end{aligned}
\end{equation}
where the 2nd and 3rd terms represent the symmetric side lobes \added{and we have omitted the arguments $\theta_\text{sim}$ and $\phi_\text{sim}$ for brevity}. The fit is then performed by minimization of the sum of squared deviations between the simulated plume data and the fit plume.

\subsubsection{Fitting bounds} Upper and lower bounds for the fitted plume parameters are given in Table~\ref{tab:fit_bounds}. These limits are inferred from the polar and azimuth plots in \cite{Jaggi2024}.%, reproduced in Figure~\ref {fig:2D_distributions}. 
The constraints are chosen in a way that the side lobes only capture direct knock-on recoils and therefore only tilt forward, away from the source of incident ions, with $\theta_\text{pl}\in[0,-\alpha_\text{in}]$ unless $\alpha_\text{in}=0$ or the deflection angle $\theta_\text{defl}=\alpha_\text{in}-\theta_\text{pl}$ exceeds the apparent maximum deflection angle of 100$^\circ$\,$\pm$\,7.5$^\circ$. Note that a model with significantly increased surface binding energies (Sec.~\ref{meth:morrissey}) will express an apparent maximum deflection angle greater than those in \cite{Jaggi2024} (Sec.~\ref{res:incSBEs}). % (Fig.~\ref {fig:2D_distributions}).
The polar tilt angles are limited to positive values (forward sputtering), with the maximum value limited by $\alpha_\text{in}$ as shown above. In the case of significant back-sputtering towards the incident beam, the $x$-values were inverted to allow the fitting of a plume away from the incidence ion.

\subsubsection{Data smoothing} When binning sputtered trajectories according to solid angle, ($\theta, \phi$) bins near $\theta=0^\circ$  (the surface normal) subtend smaller solid angles than bins near near $\theta=90^\circ$ , leading to reduced statistics in the smaller bins as compared to the larger ones. To address this imbalance, we applied a rolling mean filter to values below $\theta\leq0.3\,$rad in 0.1\,rad increments.

\begin{table}[width=.9\linewidth,cols=3,pos=htb!]
\centering
\caption{Fitting bounds tuples (min, max) \label{tab:fit_bounds}}
\begin{tabular*}{\tblwidth}{@{}l l l@{}}   
\toprule
\multicolumn{3}{l}{} \\
parameter & center lobe & side lobes \\
\midrule
$A$ & 0, 1 & 0, 0.5 \\
$\theta_\text{pl}$& 0, 45$^\circ$ & $p-7.5^\circ$, $p+7.5^\circ$ \\
$\phi_\text{pl}$& 0, 0 & 0, $\alpha_\mathrm{in}$ + 5$^\circ$ \\
$m$ & 1, 5 & 3, 10 \\
$n$ & 1, 5 & 3, 10 \\
\bottomrule
\textit{Notes} &  &  \\
\multicolumn{3}{l}{$A$ = amplitude (data normalized to 1)}\\
\multicolumn{3}{l}{$\alpha_\mathrm{in}$ = angle of incidence from surface normal}\\
\multicolumn{3}{l}{$\theta_\text{pl}$, $\phi_\text{pl}=$ plume tilt and rotation about surf. normal}\\
\multicolumn{3}{l}{ $p = \text{min}(\mathrm{abs}(\alpha_\mathrm{in}-100^\circ, 60^\circ) $}\\

\end{tabular*}

\end{table}

\subsubsection{Goodness of fit} To determine whether the data can be accurately fit with a plume made up of a single single lobe (restricted model) instead of a tri-lobe sum (unrestricted model), we applied a single-tail F-test with a 5\% confidence level. The test thereby compares the variability between the two models to the variability within the restricted model to assess whether the unrestricted model differs more than expected by chance (confidence level). First, we fit the data with a single lobe. Then, maintaining the central lobe tilt angle obtained from this fit, the side lobes were fit while allowing the amplitude of the central lobe to vary. If the fit residuals were not significantly reduced by the addition of the side lobes—based on the F-test criteria—then the restricted model (single lobe plume) was taken to be the representation of the data. %For the SPRAY data, 400 data points (n) were used to fit with 10 parameters (k) with 4 restrictions (q) based on the single lobe fit resulting in a F-statistic of $F(q,n-k-1) = F(4,389)$.

\section{Results}\label{sec:results}

\subsection{Sputter yield}

\begin{figure}
    \centering
    \includegraphics[width=\linewidth]{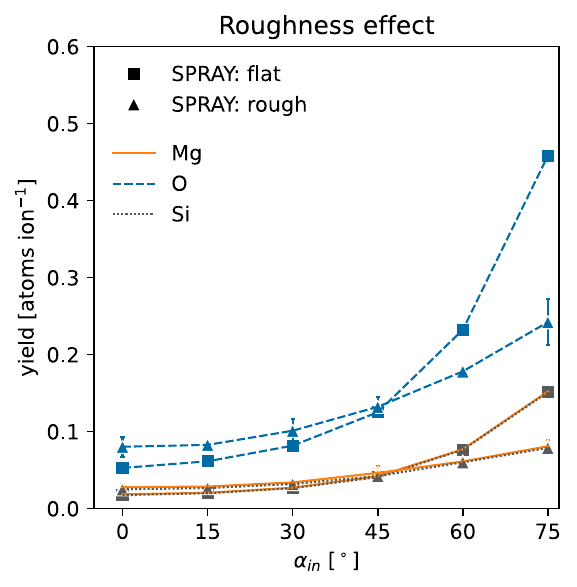}
    \includegraphics[width=\linewidth]{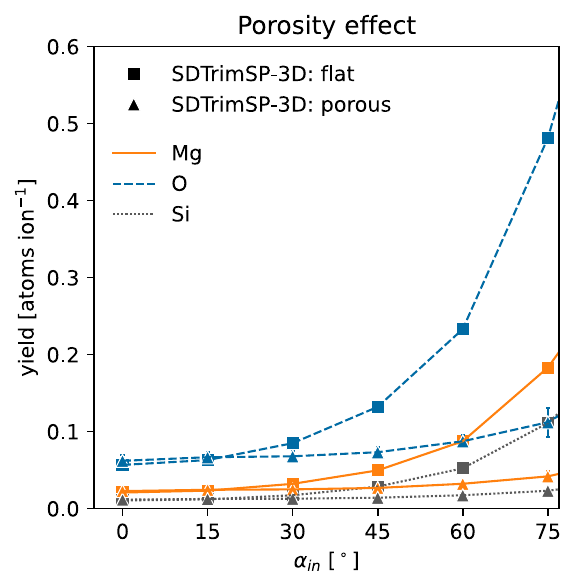}
    \caption{Elemental yield comparisons between flat (full lines) and rough surface (dashed lines). Note that Mg and Si yields of the flat surface in SPRAY overlap because the yields are stoichiometric (dynamic simulation) unlike the flat reference surface run in SDTrimSP-3D (static).}
    \label{fig:yields}
\end{figure}

We first compare surface yields of a flat target (calculated in SPRAY) with their rough (SPRAY) and porous (SDTrimSP-3D) counterparts (Fig.~\ref{fig:yields}). The composition of species sputtered from the flat surface is stoichiometric because the SDTrimSP yields that are used as SPRAY inputs represent the equilibrium state of the dynamic simulation after sufficient fluence has been attained. For this reason, SDTrimSP yields also agree with flat SPRAY compositional results (not shown). SDTrimSP-3D calculations, however, were performed statically due to an excessively large computation time required to reach equilibrium in dynamic mode and are thus not equivalent to the bulk stoichiometry. \added{}
The yield variation among the two rough SPRAY surfaces and the five porous SDTrimSP-3D structures is mostly negligible ($<10\%$). In Figure~\ref{fig:yields} below, error bars that are two standard deviations (SD) from the mean in length are generally smaller than the plotted symbols.% for all but the sputtered oxygen at incidence angles below 60$^\circ$ from the surface normal.
The relative error of an element yield ($\text{SD}_\text{el}/(\text{SD}_\text{el}+Y_\text{el})$) at any given $\alpha_\text{in}$ is less than 8$\%$.

When comparing elemental sputtering yields from a flat surface with yields from the corresponding rough or porous computation, the decrease in sputtering yield as $\alpha_\text{in}$ approaches 90$^\circ$ from the surface normal is identical between all species sputtered. At a 75$^\circ$ incidence, yields from a rough surface are about half (55\%) that of flat surface, whereas yields from a porous surface are about one fifth (22\%) of the flat surface reference (Fig.~\ref{fig:yields}). At normal and near-normal incidence, the rough surface expresses slightly larger sputtering yields. This effect is not observed in the porous case.

%\njnote{this is not true, see previous statement on where to get yields:} For a rough surface, a difference exceeding 5\% in yield was only found for $\alpha_{in}=85^\circ$. This increase is found in Si (+9\%) at the cost of Mg (-2\%) and O (-7\%). We interpret this change in the emission angle from the typical behavior of Si being sputtered more towards the surface normal, because of its large surface binding energy, causing Si contribution to not only originate from the first atomic layer. 

\subsection{Angular distribution}\label{res:angDist}

\begin{figure*}[htb!]
    \centering
    % trim: <left> <lower> <right> <upper>
    \includegraphics[trim={0cm 0cm 0cm 0cm}, clip, width=0.90\linewidth]{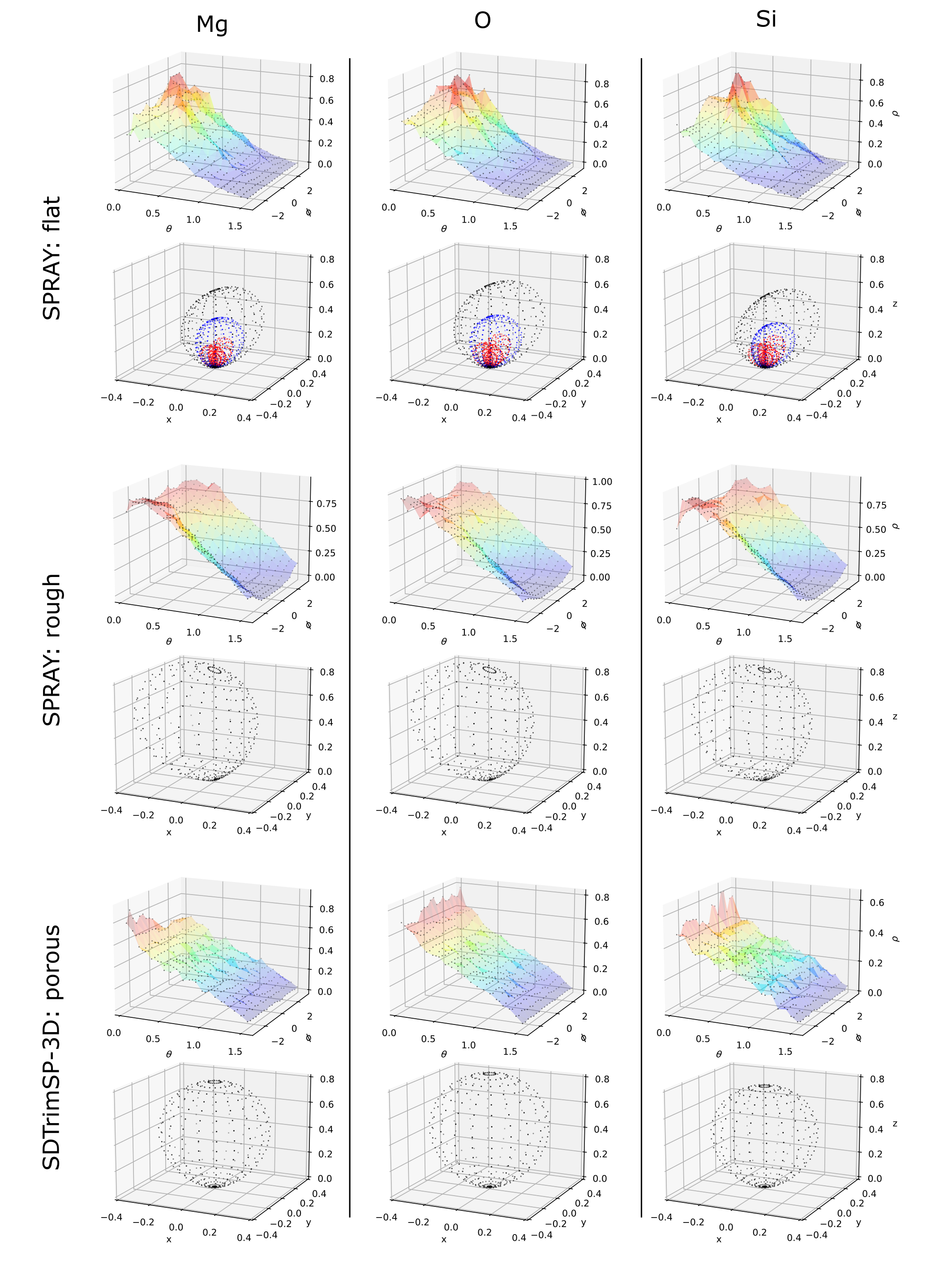}

    \caption{Angular distribution of simulated ejecta from a flat (rows 1\,\&\,2), rough (3\,\&\,4) and porous surface (5\,\&\,6) of Mg (left), O (middle) and Si (right) sputtered by 4\,keV He at an incidence angle of 80$^\circ$ from surface normal (beam originates from $-x$ direction). A component facing sideways is reflected in the ``winglets'' of the SPRAY:FLAT data and was fitted using ``side-lobes'' (red) in addition to the main lobe (blue) in the fit plume (black). \added{The colored surfaces show the ejecta distribution in $\theta-\phi$ space, highlighting the ``winglets'' and the narrow distribution in the shape of ridges leading to a peak, as well as  forward (convex) and backwards sputtering (concave).}
    }
    \label{fig:3Dangular-fit}
\end{figure*}

\subsubsection{Flat surface--triple lobe}
Employing the model developed in Section \ref{meth:fitting}, the flat surface hybrid model data \citep{jaggi_new_2023} produced in SDTrimSP and SPRAY can be reliably fitted with two symmetrical side lobes (red in Fig.~\ref{fig:3Dangular-fit}) to capture the broad single knock-on component. These side lobes are reflected in the ``winglets'' of the flat surface raw data plotted with respect to ($\theta$,$\phi$) in the surface coordinate system (Fig.~\ref{fig:3Dangular-fit}). %The angular distribution of a rough surface plotted in $\theta$--$\phi$ space is concave because of the ejecta tilting towards the ion incidence (row 4 in Fig.~\ref{fig:3Dangular-fit}).  
In the flat surface hybrid model without dissociation (static SDTrimSP-3D calculation), the single knock-ons reliably fit data produced by 45$^\circ$ and 60$^\circ$ He ion incidence, but not for $\alpha_\text{in}\geq75^\circ$. The static nature of the SDTrimSP-3D simulations is more representative of a fresh surface rather than a weathered surface. In the static simulation, the surface composition does not change with the removal of species; therefore, easily removed species dominate the yield. The larger yields lead to more surface-normal ejecta with a narrow peak that is difficult to reliably fit with the side lobes. % (e.g., ``SDtrimSP: flat, static'' in Fig.~\ref{fig:2D_distributions}).%, unlike in the dynamic case (Fig.~\ref{fig:2D_distributions}a).  \njnote{Our fitting procedure does not produce as sharp a lobe to be able to reproduce the shape.}  

\subsubsection{Rough and porous surfaces--single lobe}
The rough and porous surface structures---simulated in SPRAY and SDTrimSP-3D, respectively---both result in ejecta that are distributed nearer to the surface normal direction with a small tilt toward the ion source instead of the strongly forward-facing distribution of a flat surface (Fig.~\ref{fig:3Dangular-fit}). Unlike the flat surface results simulated in either SDTrimSP or  SDTrimSP-3D, the fit is not improved using side lobes.

\subsection{Energy distribution}
\begin{figure}[htb!]
    \centering
    \includegraphics[width=\linewidth]{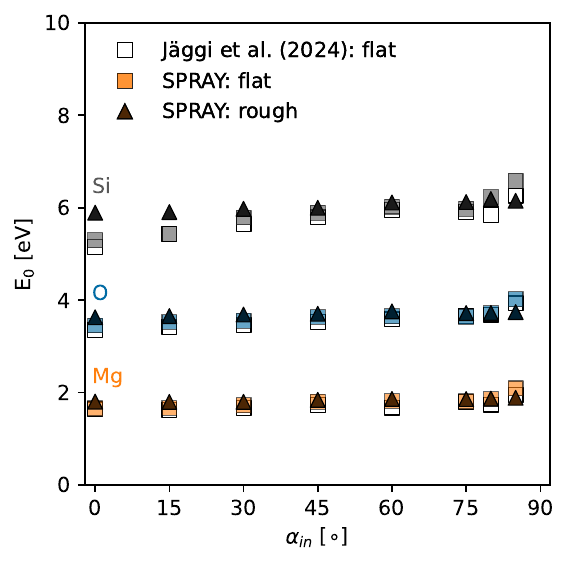}
    \includegraphics[width=\linewidth]{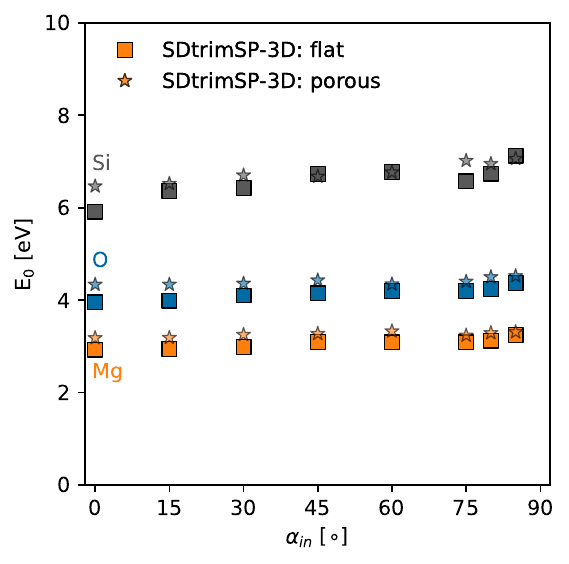}
    \caption{Characteristic energy $E_0$ of the \added{Thomson distribution fitted to the} energy distributions of particles sputtered from He-irradiated MgSiO$_3$. \added{We compare $E_0$} at incident angles $\alpha_\mathrm{in}$ from a flat \citep{Jaggi2024}, rough (SPRAY) and a porous surface (SDTrimSP-3D). The flat surface results from SPRAY are based on the input trajectories produced by the \cite{Jaggi2024} model and represent the systematic error in the energy distribution. Only at normal and near-normal incidence are the ``SPRAY: rough'' $E_0$ values significantly increased compared to the flat data. For the ``SDTrimSP-3D: porous'', the significant  $E_0$ increase is confined to Si sputtered at normal incidence.
    }
    \label{fig:char_E0}
\end{figure}

We used an adapted Thompson fit function \added{\citep{Sigmund1969,Wurz2010,Jaggi2024}} to fit the energy spectra and extracted element-specific characteristic energies ($E_{0,\text{el}}\approx \text{SBE}_\text{el}/2$). The simulated energy distribution of ejected particles from a surface is mostly unaffected by the ion incidence angle, except for the largest $\alpha_\text{in}$. This was shown for flat MgSiO$_3$ \citep[Fig.~3 in ][]{Jaggi2024} where the shift in the energy distribution was described by comparing the $E_0$ of the adapted Thompson distribution. The $E_0$ thereby describes the position of the maximum in the energy distribution. We reproduced the \cite{Jaggi2024} results by comparing the $E_0$ values from the flat surface SPRAY results with the SDTrimSP energies (Fig~\ref{fig:char_E0}). The two should be in agreement, since the SDTrimSP trajectories are used as SPRAY inputs; however, there are small deviations in the $E_0$, which we attribute to statistical outliers. Note that the SDTrimSP-3D simulations are static and without dissociation of compounds; therefore, the binding energies are increased, although still comparable to the model applied in the flat and rough surface simulations (Sec.~\ref{meth:SDTrimSP}\,\&\,\ref{meth:SPRAY}). No direct comparison between SPRAY and SDTrimSP-3D can be drawn, but the trend of an overall increased $E_0$ for the rough and porous surfaces is the same when comparing to the respective flat surface SPRAY and SDTrimSP-3D simulations. 

For the rough surface data obtained with SPRAY, the $E_0$ is nearly constant as a function of $\alpha_\text{in}$, with a slight increase  as $\alpha_\text{in}$ approaches zero (Fig.~\ref{fig:char_E0}). Compared to the flat surface, this leads to a significant increase in the characteristic energy at normal incidence. The increases in this case are 0.14 eV for Mg (+8.8\%), 0.58 eV for Si (+11\%), and 0.18 eV for O (+5.3\%). At highly grazing incidence angles ($\alpha_\mathrm{in}\geq85^\circ$ ) the rough-surface characteristic energy does not follow the same trend as that of the flat surface, such that it is lower for the rough surface than the flat one.

The SDTrimSP-3D simulations comparing flat to porous surfaces express similar trends (Fig.~\ref{fig:char_E0}). The $E_0$ of Mg and O from porous samples are overall elevated compared to their flat surface counterparts. The $E_0$ of Si expresses more variability, especially for cases where the SBE and BBE are combined (Jäggi et al. in Fig.~\ref{fig:char_E0}). Unlike the SPRAY results, SDTrimSP-3D shows no systematic discrepancy in $E_0$ between the flat and porous surfaces at near-normal incidence angles. This is attributed to the spherical grains being locally flat and minimally shadowed at near-normal incidence . %At shallow incidence angles the porous surface $E_0$ shows the best agreement with that of the flat surface, suggesting a similar topographic dependency observed in the SPRAY flat vs. rough comparison.

\subsection{Increased binding energies}\label{res:incSBEs}
\begin{figure}[htb!]
    \centering
    \includegraphics[width=\linewidth]{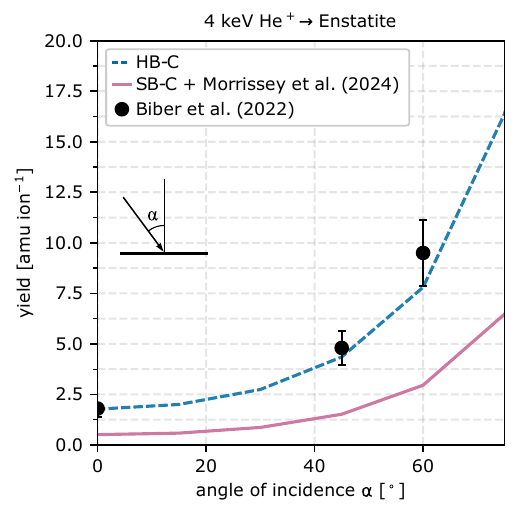}
    \includegraphics[width=\linewidth]{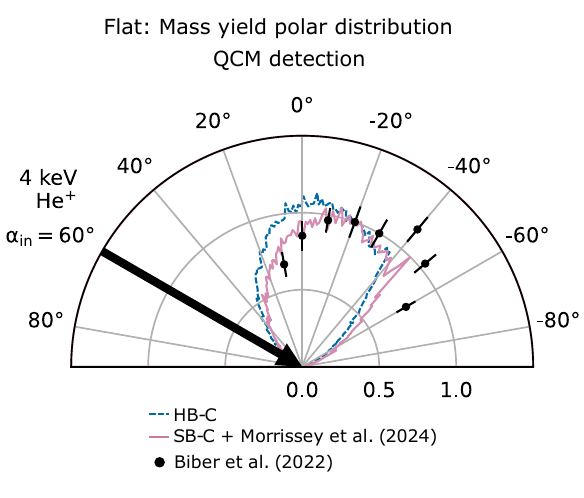}
    \caption{SDTrimSP model yield and angular distribution results compared to experimental data by \cite{Biber2022}. The compound model (-C) is used to obtain near-ideal mineral densities. The binding energies from \cite{morrissey_solar_2024} (M24) were used to increase the system's surface binding energies (SB case). %or bulk binding energies (HB case) respectively. 
    Abbreviations: HB~--~tabulated enthalpy of formation as bulk binding energy and enthalpy of sublimation as surface binding energies; %HB+M24~--~Binding energies from M24 minus the enthalpy of sublimation as bulk binding energies and enthalpy of sublimation as surface binding energies;
    SB+M24~--~Binding energies from M24 as element surface binding energies.}
    \label{fig:FLATsbbxVSsbxmYandDist}
\end{figure}

So far, we have exclusively used the \cite{jaggi_new_2023} compound model that reproduces the flat surface MgSiO$_3$ experimental sputter yields exceptionally well. When using increased SBEs based on the $E_b$ values reported in \cite{morrissey_solar_2024}, we find reduced sputter yields for a flat surface, but the angular tilt away from the incident ion direction increases (Fig.~\ref{fig:FLATsbbxVSsbxmYandDist}).

\begin{figure*}[htb!]
    \centering
    \includegraphics[width=0.9\linewidth]{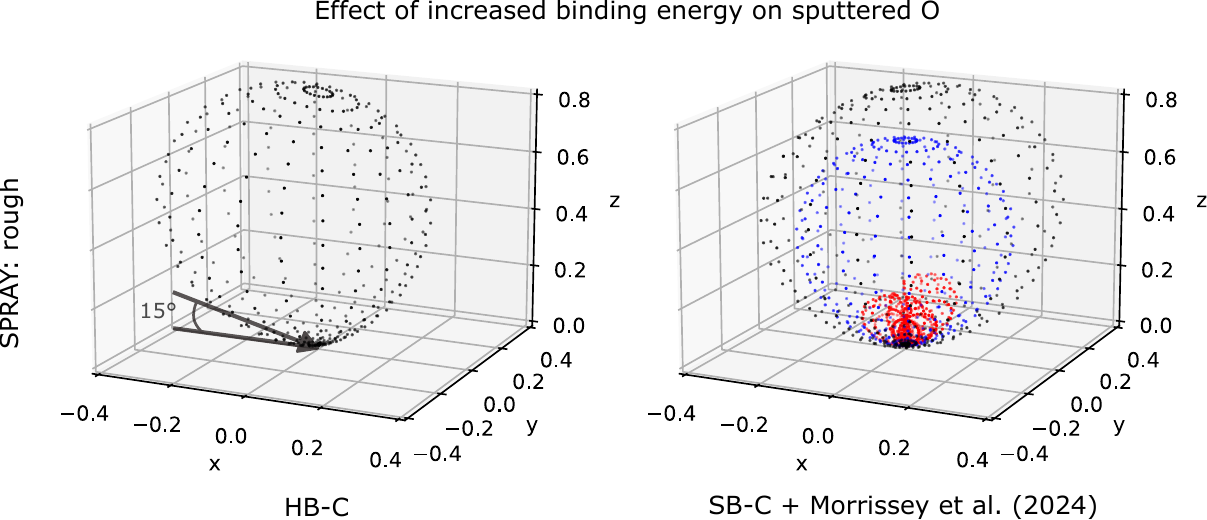}
    \caption{Comparison of O sputter ejecta distributions sputtered from a rough surface (SPRAY) by 4~keV He at an incidence angle of 75$^\circ$ from the surface normal (black arrow) resulting from (left) the \cite{jaggi_new_2023} hybrid binding energy (HB) model and from (right) a surface binding energy (SB) model with increased binding energies from \cite{morrissey_solar_2024}, fitted by a central plume (blue) and two forward-facing ``winglet'' plumes (red).}
    \label{fig:hi_SBE_3D_fit}
\end{figure*}

\begin{figure}[htb!]
    \centering
    \includegraphics[width=\linewidth]{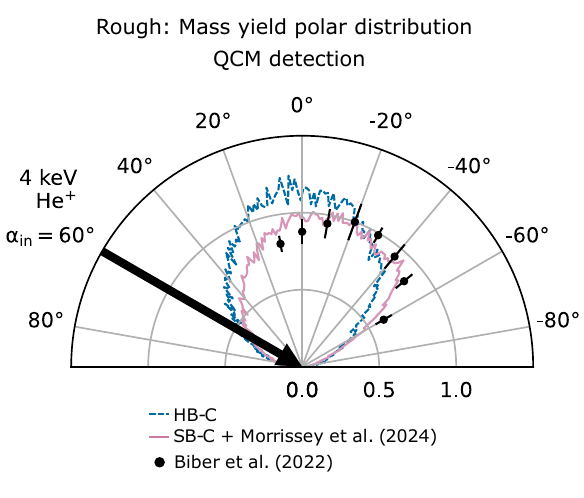}
    \caption{Comparison of mass yield (amu) sputter ejecta distributions sputtered by 4~keV He at an incidence angle ($\alpha_\text{in}$) of 60$^\circ$ resulting from the \cite{jaggi_new_2023} hybrid binding energy model (HB-C, blue dashed lines) with the surface binding using increased binding energies (SB-C) based on \cite{morrissey_solar_2024}.}
    \label{fig:ROUGHsbxmDist}
\end{figure}

To investigate the agreement with the \cite{Biber2022} MgSiO$_3$ pellet laboratory data, we performed rough surface SPRAY simulations using the pellet AFM images as topography together with the increased SBE model. The resulting tilt angles of the plume fitted to the ejecta for O, Mg, and Si are 5.6$^\circ$, 7.2$^\circ$, and 11$^\circ$ at 45$^\circ$ incidence and 3.5$^\circ$, 4.5$^\circ$, and 12$^\circ$ at 60$^\circ$ incidence, respectively. This is still lower than the 21$^\circ$--25$^\circ$ angular range of the ejected observed in the rough pellet experiment, but unlike the other binding energy models, the tilt is away from the incident ion source direction. This difference is visible even in the least affected, but yield-defining, O distribution (Fig.~\ref{fig:hi_SBE_3D_fit}). This then results in an unprecedented agreement between the QCM plume  slice measured by \cite{Biber2020} and the simulated mass yield (Fig.~\ref{fig:ROUGHsbxmDist}).  

\section{Discussion}

\subsection{Reliability of reference data} \label{dis:datareliability}
The MgSiO$_3$ reference data for yields and angular distributions from \cite{Biber2022} which we use to benchmark our simulations comes with one major caveat that complicates direct comparisons. The available angular distribution data does not resolve the type of sputtered atoms; therefore, we are not confident that the mass-change data obtained at the catcher QCM  represents each individual species sputtered into the detector solid angle. %Unlike a mass change measurement taken from a thin film-covered QCM, which encompasses all the material lost from the target, the catcher QCM only registers the mass of the species deposited. 
If there is a significant difference in the sticking probability of the ejected species, the measured angular distribution would instead represent only a subset of the ejecta. For example, if Si exhibits a significantly larger sticking probability than both O and Mg and also exhibits a stronger forward tilt as seen in the dynamic flat surface simulations (Figure~\ref{fig:1Dvs3D}), then the mass yields obtained in \cite{Biber2022} could be dominated by Si, overestimating the plume tilt of all sputtered particles. %\njnote{Fritz: Estimate the effect of element-dependent sticking in the reference QCM data.}

\subsection{Sputter yields and energy distribution}
The behavior of the yield as a function of incidence angle for rough and porous surfaces was already discussed in \citep{Cupak2021,szabo2022analytical}, \added{and most recently for flat and rough regolith samples in \cite{brotzner_solar_2025}.}. For rough surfaces, it was found that a low degree of surface roughness leads to an increase of the sputter yield at near-normal global incidence because the local incidence angles are near-normal as well. In high roughness cases, normal global incidence results in large local incidence angles and significant forward scattering into the surface such that the yield is significantly reduced through shadowing and redeposition. In the case of 80\% porosity, shadowing and redeposition dominate the yield, and no increase in yield is observed at normal incidence. 

The simulated energy distribution as a function of incidence angle follows the trend of the yield.  The initially elevated but ultimately flat distribution of $E_0$ for rough surfaces simulated in SPRAY and a lack of enhanced $E_0$ at $\alpha_\mathrm{in}>80^\circ$ is explained by the distribution of local incidence angles. Analogous to the sputter yield \cite[e.g., ][]{Gauthier1990,Roth1991,Kustner1999,Biber2022}, at normal incidence on a rough surface, an ion will impact locally at non-normal incidence angles. For rough surfaces, local surface normals diverge from the global normal such that normal global incidence results in larger local incidence angles. And conversely, at large global incidence angles, the rough surface yield is described by locally normal incidence angles. Both energy and yield are thus elevated at normal incidence, unhindered by the low degree of surface roughness \citep{Cupak2021,szabo2022analytical}, but reduced at large $\alpha_\text{in}$. We interpret the overall slightly higher $E_0$ of rough and porous surfaces compared to their flat surface counterparts to result from a small increase in the amount of slow ejecta being re-deposited onto the surface.  

\subsection{Angular distribution} \label{dis:angular}
We find that the rough surface SPRAY simulations using the \cite{jaggi_new_2023} binding model do not reproduce the angular distribution of a rough MgSiO$_3$ surface (Fig.~\ref{fig:ROUGHsbxmDist}). The best agreement with the experimental angular distribution data of a flat and rough surface \citep{Biber2022} is instead reached when using the \cite{morrissey_solar_2024} binding energies (Sec.~\ref{meth:morrissey}) at the cost of a 3--4 fold underestimate of the corresponding mass yields (Fig.~\ref{fig:FLATsbbxVSsbxmYandDist}). Moreover, for rough surfaces the ejecta tilt is forward-facing, unlike for binding energy models that reproduce sputter yields, but the tilt is still below that of the \cite{Biber2022} experimental data. We are therefore left with the conundrum that we can either reproduce the experimental sputter yields or have more realistic angular distributions, but not both. 

For porous surfaces, there is only limited angular distribution data available \citep[][wherein the porosity/morphology changes dynamically throughout the experiment]{bu_absolute_2025}; however, the work on simulating backscattered ions from regolith-like structures \citep{Szabo2022c} indicates that there is more backscattering from regolith. %This behavior may be exacerbated in the setup described in \cite{Szabo2022c} because they not only modeled smooth spheres, but also macroscopically rough particles that exceed the voxel-level roughness of a ``smooth'' 3D sphere, combining porosity and roughness effects. %\njnote{@Paul: Can you rewrite this and clarify how the spheres alone did not do the job. If I remember correctly, the backscattering by SDTrimSP-3D when using spheres is underestimated.} \psnote{The smooth spheres actually do a good job, within the limited range of rough particles that we explored, there was only a small difference.}

%We conclude, independent of the binding energy model used, that in simulations, roughness has a similar effect to porosity, causing backscattering of sputtered atoms. This may be erroneous, because based on the experimental data of \cite{Biber2022} and the \cite{Szabo2022c}, we would instead expect a small, but prominent forward tilt for the rough surface and backscattering of both ions and sputtered ions in porous surfaces.

%Overall, the near-normal distribution of the single-lobe fit to the ejecta of rough and porous samples suggests that the contribution of direct knock-on recoils to the ejecta is negligible. We would therefore expect the energy distribution of the ejecta to not be affected by direct knock-on recoils either.

%% COMMENT ON MORRISSEY STATIC SIMULATIONS
%In the case of the binding energies in \cite{morrissey_solar_2024} the authors predict a 0.2--0.3 amu~ion$^{-1}$ yield based on static irradiation simulations on anorthite and albite and compare this to the 0.2--0.3 amu~ion$^{-1}$ yield obtained by \citep{Szabo2018} H onto wollastonite. The latter however is an equilibrium yield, obtained after the surface reaches a steady state based on the preferential removal of of low-mass, low-binding energy species and ion implantation, which further reduces the mass yield per impinging ion. To compare static results with experimental data, the mass yields at the beginning of the irradiation process would have to be used instead. 

\subsection{Outstanding validation by experimental data}
\added{This work touches on several open questions that are yet to be addressed through experimental work. The composition of the sputtered ejecta from silicate minerals is yet unknown, and would inform on the appropriateness of using dynamic over static simulations, and vice versa. Furthermore, measuring the speed of the ejected particles would help validate the increased SBEs predicted by MD, and tell us if SDTrimSP accurately reproduces the angular distribution given the experiment-validated SBEs. Lastly, we briefly mentioned the existing laboratory angular distribution data, which exists for solid and regolith-like Cu \citep{bu_absolute_2024,bu_absolute_2025}. There, the angular distribution ejecta created by 20~keV Kr$^+$ could be well reproduced by a multiscale MD-BCA-MC approach, using the default enthalpy of sublimation of Cu for obtaining ejecta angular distributions in SDtrimSP \citep{verkercke_theoretical_2026}.}

\added{The difference to the Kr on Cu powder experiments are twofold: Firstly, We have shown in this work for the silicate mineral enstatite, that increased binding energies of the compound model of \cite{jaggi_new_2023} is incapable of reproducing laboratory angular distribution data whereas the MD-sourced SBEs akin to those presented in \cite{morrissey_solar_2024} are capable to accurately predict the plume tilt. Secondly, unlike for 20~keV Kr$^+$ on Cu, nuclear stopping dominates over electronic stopping about 10:1 (according to SRIM-2013), while the same ratio is about 1:2 for 4 keV He on MgSiO$_3$. The dominant electronic stopping for He will cause very different paths of the ion in the solid and a different sputtering behavior. This difference was shown in the close-to normal angular distribution of SiO$_2$-sourced ejecta produced by 20~keV Kr, when compared to H, and a lesser degree He, ions \citep{clouter-gergen_investigating_2026}.} 

\added{We have shown that the angular distribution of ejecta produced by 4~keV He irradiated enstatite is quite sensitive to an increase in SBEs, with the capability of reducing ion-incidence facing, backward sputtering in the emission from rough and porous enstatite surfaces. Since plagioclase and enstatites are the major regolith minerals on the Moon and possibly Mercury \citep{Heiken1991,McCoy2018}, their flat-surface ejecta behavior informs on the resulting porous regolith sputter ejecta distribution. The proposed increased binding energies \citep[e.g.,][]{morrissey_solar_2024} would suggest a near-normal distribution of ejecta being released from regolith. To verify if such a normal distribution over porous regolith holds true, experiments akin to those in \cite{bu_absolute_2025} are needed.}

\section{Conclusions}
As presented in this manuscript, there are challenges with BCA simulations when attempts are made to match experimental data for sputter yields and angular distributions simultaneously. Sputter models that accurately reproduce yields differ regarding the simulated ejecta from rough and porous surfaces, which have an enhanced angular distribution towards the surface normal. We furthermore investigate the peak energies of sputtered atom energy distributions and find that peak energies are slightly increased compared to simulations on a flat surface. For a rough surface, we are confident that the tilt of the distribution of simulated ejecta using commonly applied surface binding energies \citep[discussed in ][]{morrissey_establishing_2023,jaggi_new_2023} do not reproduce laboratory observations. Instead, we have demonstrated that increasing SBEs, akin to the utilizing those proposed by \cite{morrissey_solar_2024} for plagioclase, provides a match to the forward tilt in the angular distribution of products sputtered from a rough surface experiments, as shown in \cite{Biber2022}, while underestimating the absolute sputtering yields. Currently, no existing model can reproduce both angular distributions and sputter yields \added{on the silicate enstatite} at the same time, representing a significant challenge for consistent modeling of the sputtering process from a realistic planetary-type surface. \added{This leaves us to wonder if}: 

\begin{itemize}
    \item SDTrimSP is incapable of accurately reproducing angular distributions of sputtered particles \added{from enstatite, and possibly all oxide minerals,} and the \cite{jaggi_new_2023} compound model binding energies \added{are appropriate to approximate the total yield and its distribution}.
    \item The binding energies from \cite{morrissey_solar_2024} are in the correct order of magnitude, but yields \added{from enstatite (and likely other silicates)} are enhanced by other processes, which are not reproduced in SDTrimSP.    
\end{itemize}

Processes that could enhance sputtering are, e.g., the sputtering of clusters \citep{wurz_interplay_2025}. In the work of \cite{Dukes2015} the secondary ion mass spectrometry calibration data of synthetic Na-bearing forsterite (MgSiO$_3$) have shown that in addition to Na$^+$ and Si$^+$, Mg is present in the form of both Mg$^+$ and MgO$^+$. The fraction of MgO$^+$ was thereby only 4\% of the counts of the Mg$^+$ signal. It is not possible to directly infer the amount of sputtered neutrals from the secondary ion population, but the presence of MgO$^+$ suggests that molecules will be sputtered intact from a planetary surface. Experiments where the angular distribution of neutrals, ions, and molecules/clusters are necessary to understand if the integrated mass yield \citep[e.g.][]{Biber2022} is affected by any of the three populations. %If we make the unlikely assumption that the secondary ions reflect the composition of the ejecta sputtered as neutrals, then we would not expect a significant resulting increase in mass yield. A 1.6 times higher mass in 4\% of the sputtered Mg would result in a total mass yield difference of 6\%. For the Morrissey binding energy model, we would need at least a 200\% yield increase to reproduce the experimental data, however. 

%This behavior is attributed to the geometric effects that cause a bias toward a normal distribution of the ejecta caused by less shallow, local incidence angles and re-deposition of forward-sputtered particles. As a side effect, we have found that the contribution of the single knock-on component---if it is not an artifact---to the formation of ejecta from rough and highly porous, regolith-like targets is negligible.

\subsection{Recommendations to exosphere modelers}
Based on the findings of this and previous works we encourage exosphere modelers to consider that: 
\begin{itemize}
    \item[a)] \textit{Sputter yields} produced with the \cite{jaggi_new_2023} compound model give a good approximation of mineral sputtering yields.
    \item[b)] \textit{Angular distributions} are best reproduced using increased surface binding energies in the magnitude proposed in \cite{morrissey_solar_2024}.
    \item[c)] \textit{Energy distributions} of the ejecta are still a topic of ongoing research and until there is energy distribution data of atoms sputtered from minerals, no recommendation can be made in the choice of binding model. To approximate the energy distribution, an energy distribution calculated at a 45$^\circ$ ion incidence relative to the surface normal is a good approximation for incidence angles of 0$^\circ$--80$^\circ$ 
\end{itemize}

\section*{Acknowledgements}
The authors acknowledge UVA Research Computing \href{https://rc.virginia.edu}{(https://rc.virginia.edu)} at the University of Virginia, UBELIX \href{https://www.id.unibe.ch/hpc}{(https://www.id.unibe.ch/hpc)} the HPC cluster at the University of Bern, and the Austrian Scientific Computing (ASC) \href{https://asc.ac.at/}{(https://asc.ac.at/)} infrastructure for providing computational resources and technical support that have contributed to the results reported within this publication.

% \textbf{Author contributions.} Noah Jäggi: Conceptualization, Data curation, Formal analysis, Funding acquisition, Investigation, Methodology, Project administration, Software, Validation, Visualization, Writing (original draft), and Writing (review \& editing).
% Adam K. Woodson: Conceptualization, Formal analysis, Investigation, Methodology, Software, and Writing (review \& editing)
% Paul S. Szabo: Investigation, Writing (original draft), and Writing (review \& editing)
% Johannes Brötzner: Investigation, Writing (original draft), and Writing (review \& editing)
% Friedrich Aumayr: Supervision, Resources, and Writing (review \& editing)
% Catherine M. Dukes: Project administration, Supervision, Resources, and Writing (review \& editing)

\section*{Code availability} 
SPRAY is openly available under the MIT software license at TU Wien Research Data (\hyperlink{https://doi.org/10.48436/1mc13-bc574}{doi: 10.48436/1mc13-bc574}). A license of the flat surface (1D) and 3D version of SDTrimSP can be acquired by contacting \hyperlink{mailto:sdtrimsp@ipp.mpg.de}{sdtrimsp@ipp.mpg.de} or the Max-Planck-Innovation GmbH. %Numerical codes must both be findable and accessible according to FAIR principles (\href{https://www.go-fair.org/fair-principles/}{https://www.go-fair.org/fair-principles/}). Findable means that the code can be located using a persistent identifier, such as a software heritage identifier SWHID, an astrophysics source code library identifier ASCL, or a DOI. The conditions under which the code and software used in a manuscript can be accessed must be described either on the landing page of the code or in the code archive. If the code is restricted and requires credentials to be accessed, the exact conditions under which the credentials can be obtained should be provided. For further information, see the journal's Open Science policies: \href{https://planetary-research.org/open-science}{https://planetary-research.org/open-science}.

\section*{Data availability}
The raw BCA and ray-tracing data including the input files used to run the codes are published on Zenodo (\href{https://doi.org/10.5281/zenodo.18417312}{doi:10.5281/zenodo.18417312}). %\added{\textcolor{red}{For review:} \href{https://1drv.ms/f/c/cdc284290320a319/IgCwFRz44DTKQaoemtnm8IgcAZTRVswfuJg5ViVRCgWSkAc?e=2DlO3Z}{OneDrive repository}} %Any data that were used or generated as part of the study and that are not found in the manuscript and supplement must be uploaded to a public data repository before the article is accepted for publication. The datasets and any accompanying metadata and documentation need to be made available to the reviewers during the peer-review process. If it is not feasible to provide all data, such as for a simulation with many time steps, the authors should provide the complete initialization files of the simulation so that the simulation could be later reproduced.

%The datasets and any accompanying metadata and documentation need to be made available to the reviewers during the peer-review process. When possible, the datasets should be shared from the repository where they will be ultimately published using a private link. For simplicity, the journal will also accept a private link to a shared folder on a cloud storage platform. During the peer review process, the authors should explicitly cite these datasets in their manuscript using a provisional citation and then update this citation at the time of final acceptance.

\section*{Funding} %\textcolor{red}{All sources of funding for the work must be disclosed. For work funded by a research agency, the following information must be provided: The name of the funding agency, the name of the funding program, the title of the funded project, and the grant number or funding reference.}

Financial support has been provided to N. Jäggi by the Swiss National Science Foundation Fund ``Quantification of irradiation-driven evaporation and characterization of potential sputtering on Mercury'' (P500PT\_217998) as well as the APART-USA Fellowship (A-12349) of the Austrian Academy of Sciences, and P. S. Szabo by the NASA's Solar System Research Virtual Institute (SSERVI) via the LEADER team (Grant 80NSSC24M0084). The authors further acknowledge TU Wien Bibliothek for financial support through its Open Access Funding Programme.

% \textbf{Competing interests} The authors declare that they have no known competing financial interests or personal relationships that could have appeared to influence the work reported in this paper.

% ------ Appendix and methods -----------------------------
%% The Appendices part is started with the command \appendix;
%% appendix sections are then done as normal sections
%% \appendix

% To print the credit authorship contribution details
\printcredits

%% Loading bibliography style file
%\bibliographystyle{model1-num-names}
\bibliographystyle{cas-model2-names}

% Loading bibliography database
\bibliography{bibfile}

% Biography
%\bio{}
% Here goes the biography details.
%\endbio

%\bio{pic1}
% Here goes the biography details.
%\endbio

\end{document}